\documentclass[fleqn,usenatbib]{mnras}

\usepackage{newtxtext,newtxmath}

\usepackage[T1]{fontenc}
\usepackage{ae,aecompl}
\usepackage[normalem]{ulem}

\usepackage{graphicx}	
\usepackage{amsmath}	
\usepackage{amssymb}	

\title[The spin-down of AR Sco]{A reevaluation of the proposed spin-down of the white dwarf pulsar in AR Scorpii.}

\author[S. B. Potter et al.]{
Stephen B. Potter,$^{1}$\thanks{E-mail: sbp@saao.ac.za}
and David A. H. Buckley$^{1}$
\\
$^{1}$South African Astronomical Observatory, PO Box 9, Observatory, 7935, Cape Town, South Africa\\
}

\date{Accepted XXX. Received YYY; in original form ZZZ}

\pubyear{2017}

\begin{document}
\label{firstpage}
\pagerange{\pageref{firstpage}--\pageref{lastpage}}
\maketitle

\begin{abstract}
We present high-speed optical photometric  observations, spanning $\sim$2 years, of the recently-discovered white dwarf pulsar AR Scorpii. The amplitudes of the orbital, spin and beat modulations appear to be remarkably stable and repeatable over the time span of our observations. It has been suggested that the polarized and non-polarized emission from AR Scorpii is powered by the spin-down of the white dwarf. However, we find that our new data is inconsistent with the published spin-down ephemeris. Whilst our data is consistent with a constant spin period further  observations over an extended time-base are required in order to ascertain the true spin-evolution of the white dwarf.
This may have implications for the various models put forward to explain the energetics and evolution of AR Scorpii.

\end{abstract}




\section{Introduction}

AR Scorpii (hereafter AR Sco) is a 3.56h binary system consisting of a rapidly spinning ($P_s$ = 117 s) white dwarf and an M-type  main sequence companion star. It was discovered to pulse across the electromagnetic system, from UV to radio, dominantly at the $\sim$118s beat period \citep{Marsh2016}. More recently \citep{takata2017} reported that the UV/X-ray emission observed with ({\it XMM-Newton}) also shows orbital and beat modulations. The spectral energy distribution of AR Sco is characterized by two synchrotron power law ($ S_{\nu} \propto \nu^{\alpha}$) components, with the addition of the M-star spectrum \citep{Marsh2016} and a hot multi-temperature thermal plasma ($kT \sim 1-8$ keV, \cite{takata2017}). For radio to infrared frequencies ($\nu \leq 10^{12} - 10^{13} \, \mbox{Hz}$) $\alpha \sim 1.3$, typical of self absorbed synchrotron emission. For higher frequencies ($\nu \geq \mbox{few} \times 10^{14} \, \mbox{Hz}$), from optical to X-rays, $\alpha \sim -0.2$ \citep{Marsh2016, Geng2016}. \cite{Marcote2017} and  \cite{Littlefield2017} report on high-angular-resolution radio interferometric observations and long-cadence optical observations respectively.


The optical pulsed emission was discovered to be highly linearly polarized, up to $40\%$, \citep{Buckley2017}, which led to the interpretation that AR Sco is the first white dwarf pulsar, with a spin period of $\sim$117s. Various models consisting of magnetic interactions between the two stars, accelerating relativistic electrons producing synchrotron radiation and MHD interactions have been put forward to explain the power behind the observed polarized and non-polarized emission, e.g. \cite{Marsh2016}, \cite{Buckley2017}, \cite{takata2017}, \cite{takata2018}, \cite{Geng2016} and \cite{Katz2017}.

\subsection{The white dwarf spin period} 

\cite{Marsh2016} report that the white dwarf in AR Sco is slowing down at a rate of 
$\dot \omega = -(2.86\pm0.36)\times10^{-17}$Hz s$^{-1}$.
It is suggested that the source of AR Sco's observed luminosity comes from the spin-down power of the highly magnetic ($\sim$500 MG, \cite{Buckley2017}) white dwarf, through dissipation of dipole radiation. Consequently an accurate measure of the spin-down rate is important for the understanding of the energetics and evolution of AR Sco.

We have observed AR Sco photometrically, with high-time resolution, on multiple occasions over a timespan of $\sim$2 yr, between 2015 and 2017 (see Table. 1). Over this time the total accumulated error is $\sim$ 20 seconds, using the quoted 
uncertainties for the spin and spin-down frequencies in \cite{Marsh2016}. This is significantly shorter than the white dwarf spin period of $\sim$117s and therefore there is no ambiguity in the spin cycle count over our datasets.

Over the course of the two years of our dataset (2015 to 2017) the spin-down rate will result in an accumulated shift in spin-phase of $\sim$0.06 equivalent to $\sim$7s.
Our data sets have a time resolution of $\sim$1s and, in addition, the spin and beat pulses are of high-amplitude on short time scales. 
Therefore it should be possible to confirm the spin-down rate reported by \cite{Marsh2016} with our new dataset.


In the next sections we present our new observations from the 2016 and 2017 observing seasons and the results of our period analysis combined with the 2015 observations from \cite{Marsh2016}. We show that our observations are inconsistent with the spin-down ephemeris of \cite{Marsh2016}.


\section{Observations}

\begin{figure}
    \centering
	\includegraphics[width=\columnwidth]{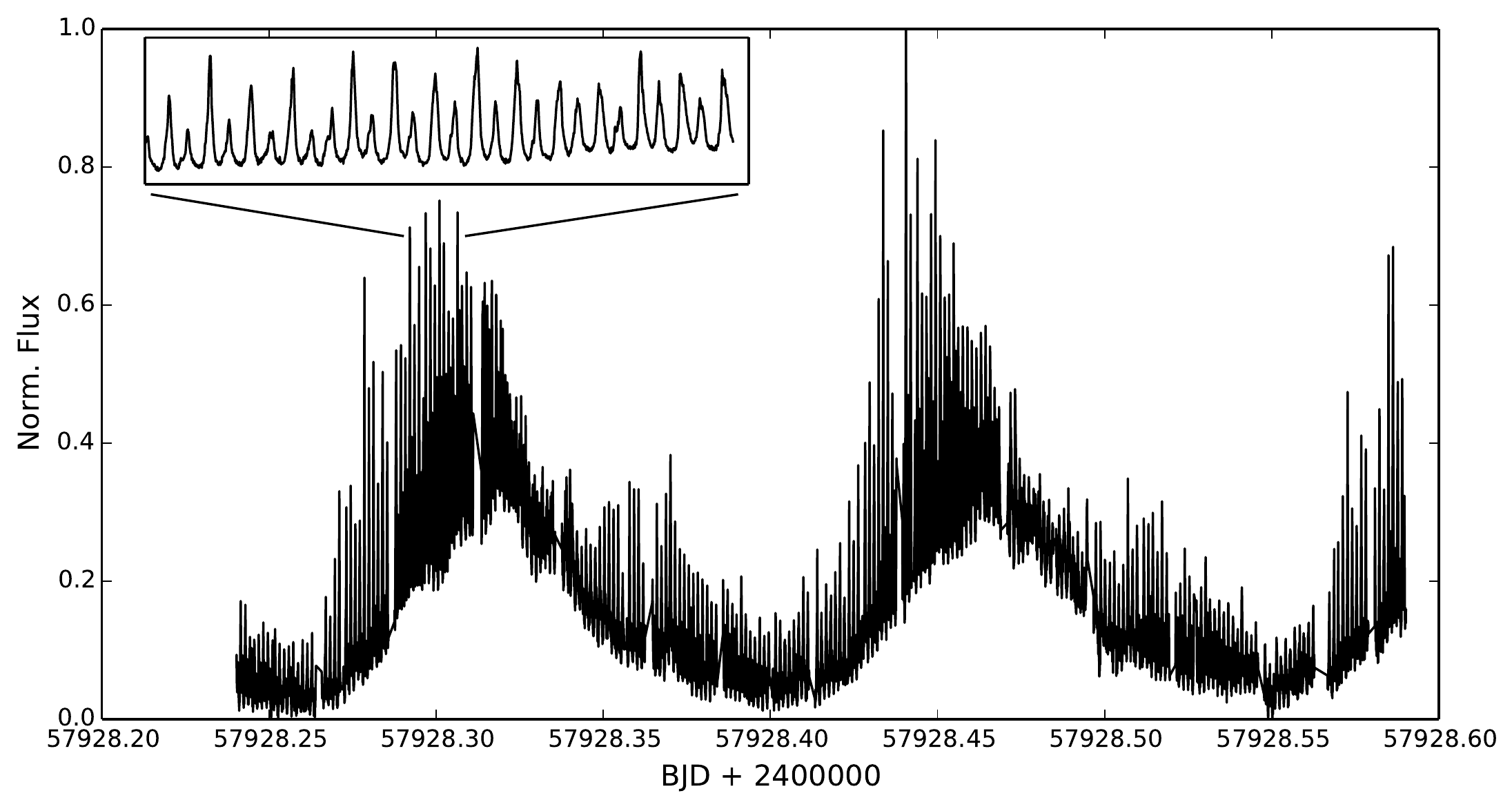}
    \caption{An example data set, specifically the $\sim$8 hours, clear filtered HIPPO observations of 23 June 2017. The inset is an expanded view of $\sim$30 minutes showing the spin and beat pulses.}
    \label{fig:EG23June2017}
\end{figure}

\begin{figure}
    \centering
	\includegraphics[width=\columnwidth]{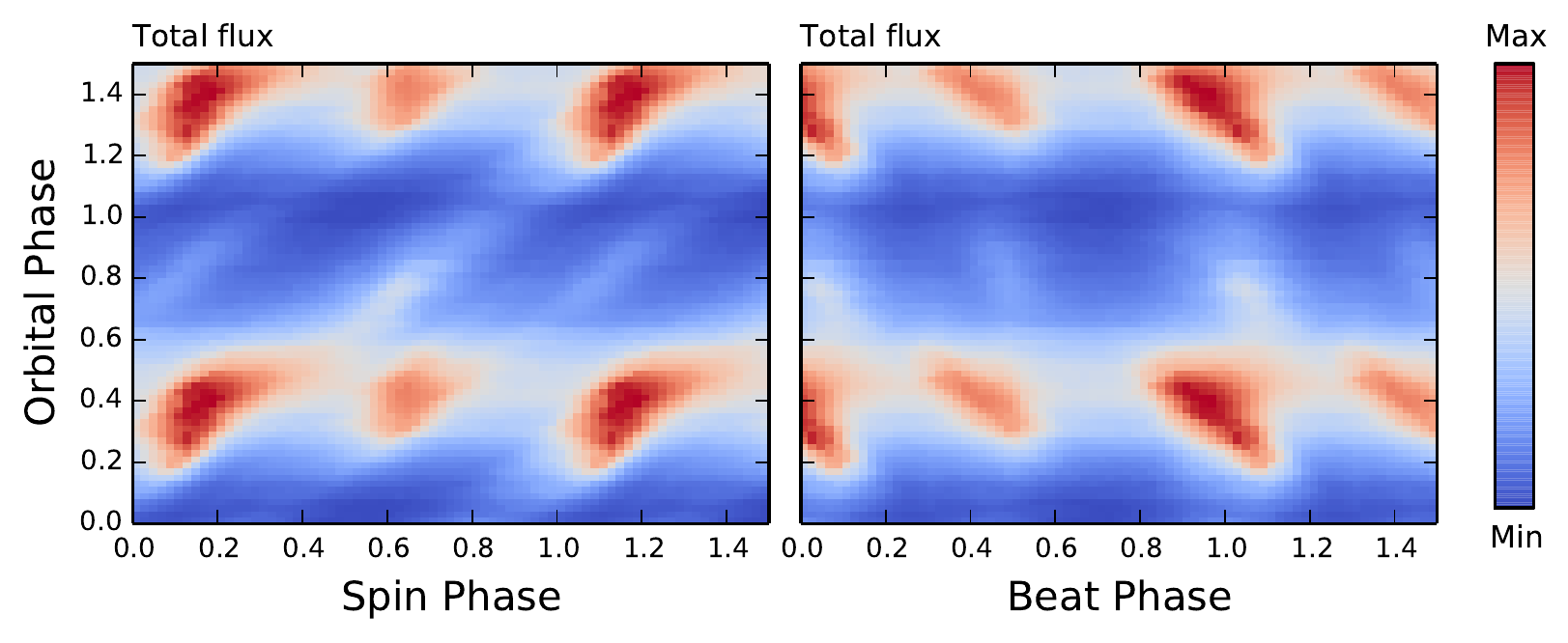}
    \caption{Left and right panels are the spin and beat folded light curves as a function of orbital phase. Photometry is normalised and the brightness is indicated by the colour bar.}
    \label{fig:example_figure}
\end{figure}

Table 1 shows a log of all the high speed photometry observations of AR Sco included in this study, which includes the original 2015 ULTRACAM photometry reported in \cite{Marsh2016} plus observations obtained by us at the South African Astronomical Observatory (SAAO) in 2016 and 2017. Observations were made with the \textit{HI-speed Photo-POlarimeter} (HIPPO; \cite{potter2010}) on the SAAO 1.9-m telescope.

The HIPPO was operated in its simultaneous linear and circular polarimetry and photometry mode (all-Stokes). Unfiltered (defined as `clear') observations (3500-9000 \AA, defined by the response of the two RCA31034A GaAs photomultiplier tubes), and filtered observations were undertaken, the latter using OG570 and I filters. The analysis of the polarimetric observations will be the subject of a separate publication.

Photometric calibrations were not carried out; photometry is given as total counts. Background sky measurements were taken at frequent intervals during the observations. 

All of our observations were synchronized to GPS to better than a millisecond. Given the high-speed nature of the instruments, their timing accuracy has been verified through feeding GPS pulsed LED light fed through the instruments.  

We corrected all times for the light travel time to the barycentre of the Solar system, converted to the barycentric dynamical time (TDB) system as Barycentric Julian Date (BJD; see \citep{Eastman2010}, for achieving accurate absolute times and time standards). By doing this we have removed any timing systematics, particularly due to the unpredictable accumulation of leap seconds with UTC, and effects due to the influence of, primarily Jupiter and Saturn, when heliocentric corrections only are applied. HIPPO data reduction then proceeded as outlined in \citep{potter2010}. 

\section{The photometry}

Fig. 1 shows our longest observation, specifically the clear-filter HIPPO observations taken on the night of 2017 June 23/24 for a total of $\sim$8 hours. Clearly seen is the high amplitude orbital modulation covering $\sim$ 2.25 cycles as well as the strong beat and spin pulses at $\sim$2 minutes. An expanded view of part of the light curve can be seen in the inset revealing the spin/beat pulses in more detail. This is consistent with the observations reported by \cite{Marsh2016} and \cite{Buckley2017}.

Fig.2 shows the details of the spin and beat modulations as a function of orbital phase, in the form of a 2D colour coded image, also known as a dynamic pulse profile. The figure was constructed by phase-fold binning on the orbital ephemeris of \cite{Marsh2016} and the spin/beat ephemeris derived in section 4. The best signal-to-noise HIPPO clear filtered 2016 and 2017 observations were used to increase the signal-to-noise. The photometric orbital, spin and beat modulations are remarkably stable and repeatable over all our data sets. Individual data sets show the same orbital/spin/beat 2D image albeit at lower signal-to-noise.

Both 2D images show that the double-peaked spin and beat pulses evolve in amplitude over the orbital cycle peaking at $\sim$0.4-0.5 in orbital phase and significantly reduced at orbital phase $\sim$0. There also appears to be a second set of double pulses (spin and beat) between orbital phases $\sim$0.6-1.0. The spin and beat pulses also appear not to be stable in phase, i.e. the spin and beat pulses appear to drift later and earlier respectively as a function of orbital phase, giving the diagonal appearance. The ``slopes" of the diagonal pulses in the spin/beat-orbit phase-space is consistent with cross ``contamination" between the spin and beat frequencies.

\begin{table}
	\centering
	\caption{Table of observations. Observations were made with the HIgh-speed-Photo-Polarimeter \citep{potter2010} on the SAAO 1.9m telescope with a cadence of 1s. ULTRACAM observations are from \citep{Marsh2016} and have a cadence of 1.3s}
	\label{tab:example_table}
	\begin{tabular}{lccrr} 
		\hline
		Date & No.Hours & Filter(s) & Instrument\\
		\hline
		24 Jun 2015 & 2.75 & g & ULTRACAM\\
14 Mar 2016 & 0.57 & OG570, clear & HIPPO\\
		15 Mar 2016 & 1.68 & OG570, clear & HIPPO\\
        14 May 2016 & 5.6 & clear & HIPPO\\
		15 May 2016 & 7.8 & OG570 & HIPPO\\
		16 May 2016 & 6.28 & I & HIPPO\\
		25 May 2016 & 4.7 & OG570, clear & HIPPO\\
		26 May 2016 & 7.37 & OG570, clear & HIPPO\\
		27 May 2016 & 6.82 & OG570, clear & HIPPO\\
		28 May 2016 & 5.98 & OG570, clear & HIPPO\\
		22 Mar 2017 & 1.68 & OG570 & HIPPO\\
		27 Mar 2017 & 4.03 & clear & HIPPO\\
		28 Mar 2017 & 4.07 & clear & HIPPO\\
		23 Jun 2017 & 8 & clear & HIPPO\\
\hline	
	\end{tabular}
\end{table}

\begin{figure}
    \centering
	\includegraphics[width=\columnwidth]{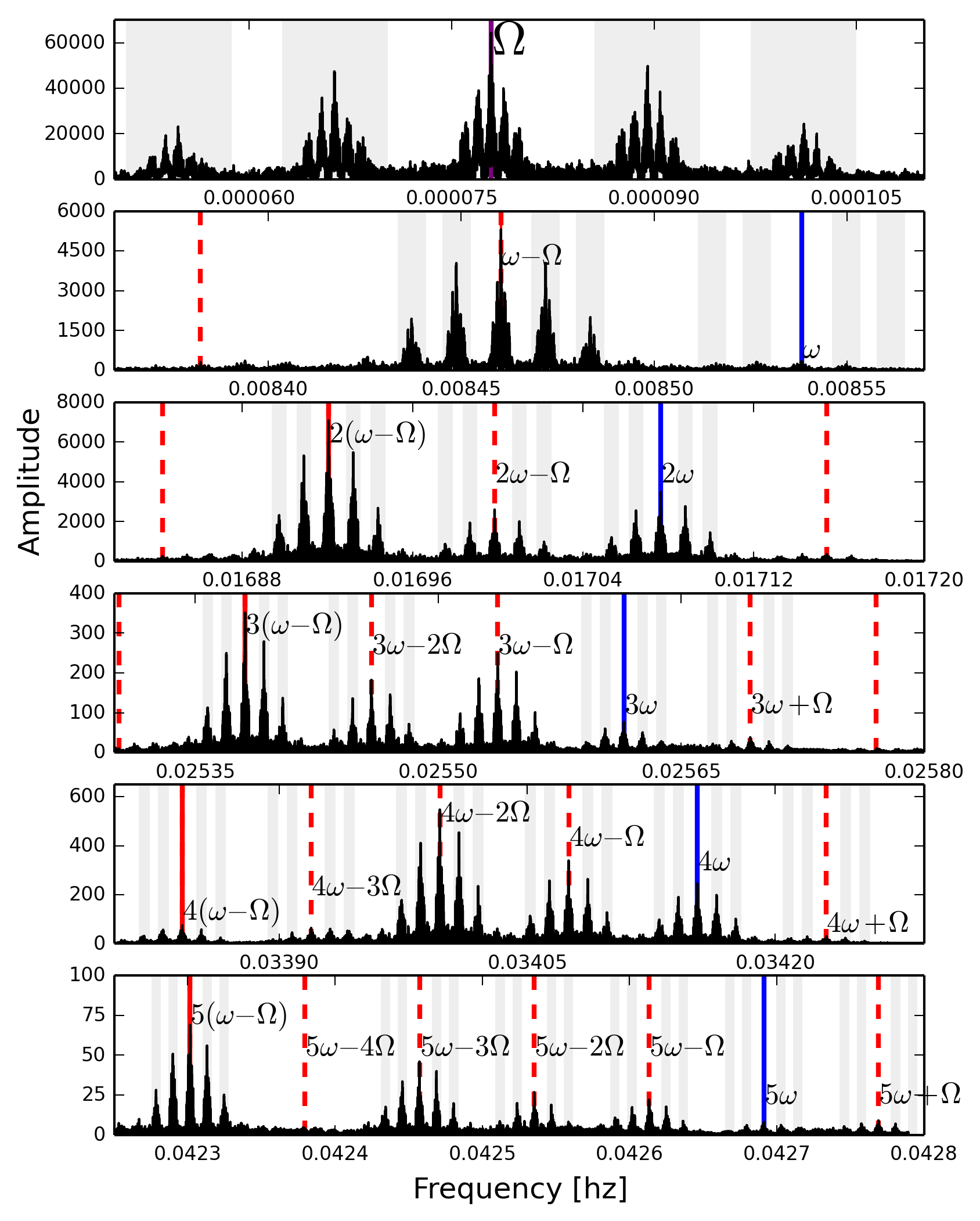}
    \caption{Amplitude spectrum of all of the photometry in Table 1. Top to bottom shows amplitude spectra centered on regions around the orbital frequency and their sidebands of the spin and orbital frequencies. Solid blue and red lines indicate the spin and beat frequencies respectively and their harmonics. Dashed red lines indicate other spin/orbit frequency sidebands. Vertical grey bars indicate one day aliases. $\sim$15 day aliases are also visible, particularly in the top panel. }
    \label{fig:example_figure}
\end{figure}

\begin{figure}
	\includegraphics[width=\columnwidth]{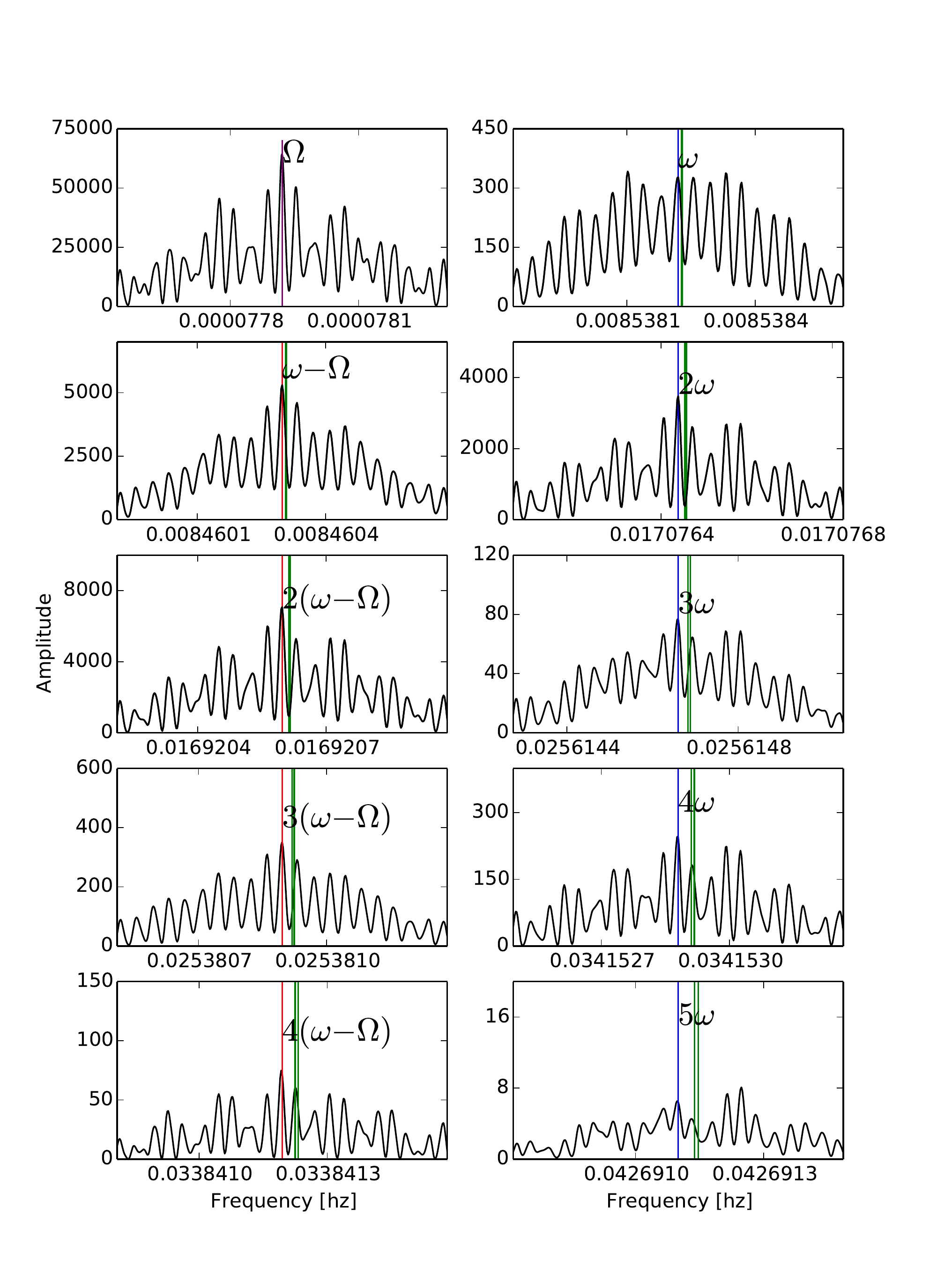}
    \caption{Magnified views of the amplitude spectrum. Left and right columns show spectral regions around the orbital/beat harmonics and spin harmonics respectively. In each case the broad spectral width corresponds to the $\sim$15-20 day aliasing. The finer peaks correspond to the $\sim$1 year aliasing. Vertical green lines correspond to the spin and beat periods from \citep{Marsh2016} adjusted for the proposed spin down rate to 2015 and 2017. Solid blue and red vertical lines correspond to our best solutions for the spin and beat frequencies.}
    \label{fig:example_figure}
\end{figure}

\section{Fourier Analysis}

\begin{figure}
	\includegraphics[width=\columnwidth]{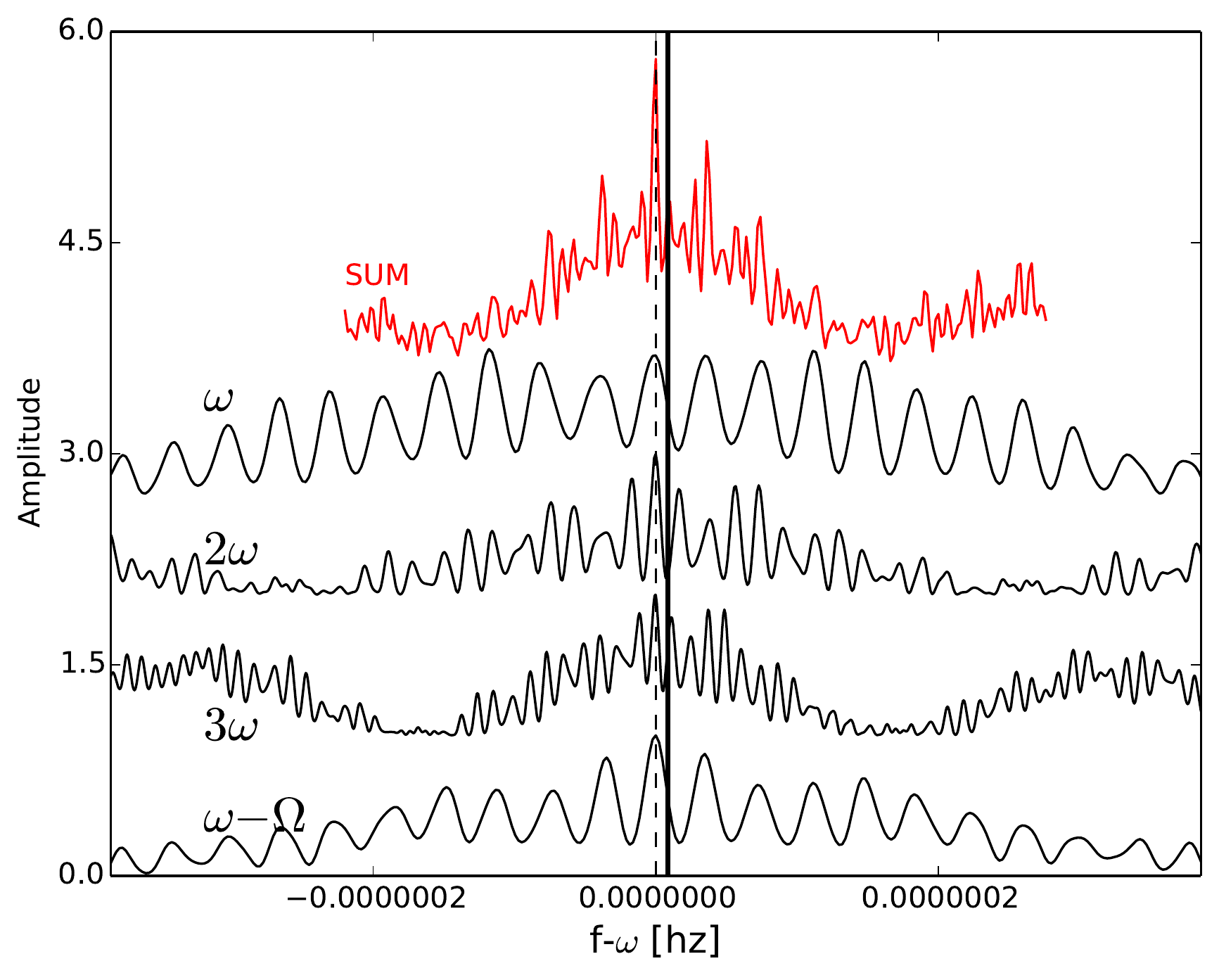}
    \caption{Black curves are magnified views centered on a sample of spin and beat harmonics re-scaled in frequency space with respect to the spin fundamental. e.g. the frequencies centered on $2\omega$ were divided by 2. Once rescaled the black curves (including others identified in Fig. 3 but not plotted here) were summed to produce the red curve. All spectra have been vertically displaced to aid visualization. The vertical solid line represents the location of the spin and beat harmonics assuming the spin frequency (corrected for spin down to 2015 and 2017) using the \citep{Marsh2016} spin ephemeris. The vertical black dashed line indicates the location of the spin and beat harmonics using the center of the tallest peak in the summed red spectrum. }
    \label{fig:example_figure}
\end{figure}


\begin{figure}
	\includegraphics[width=\columnwidth]{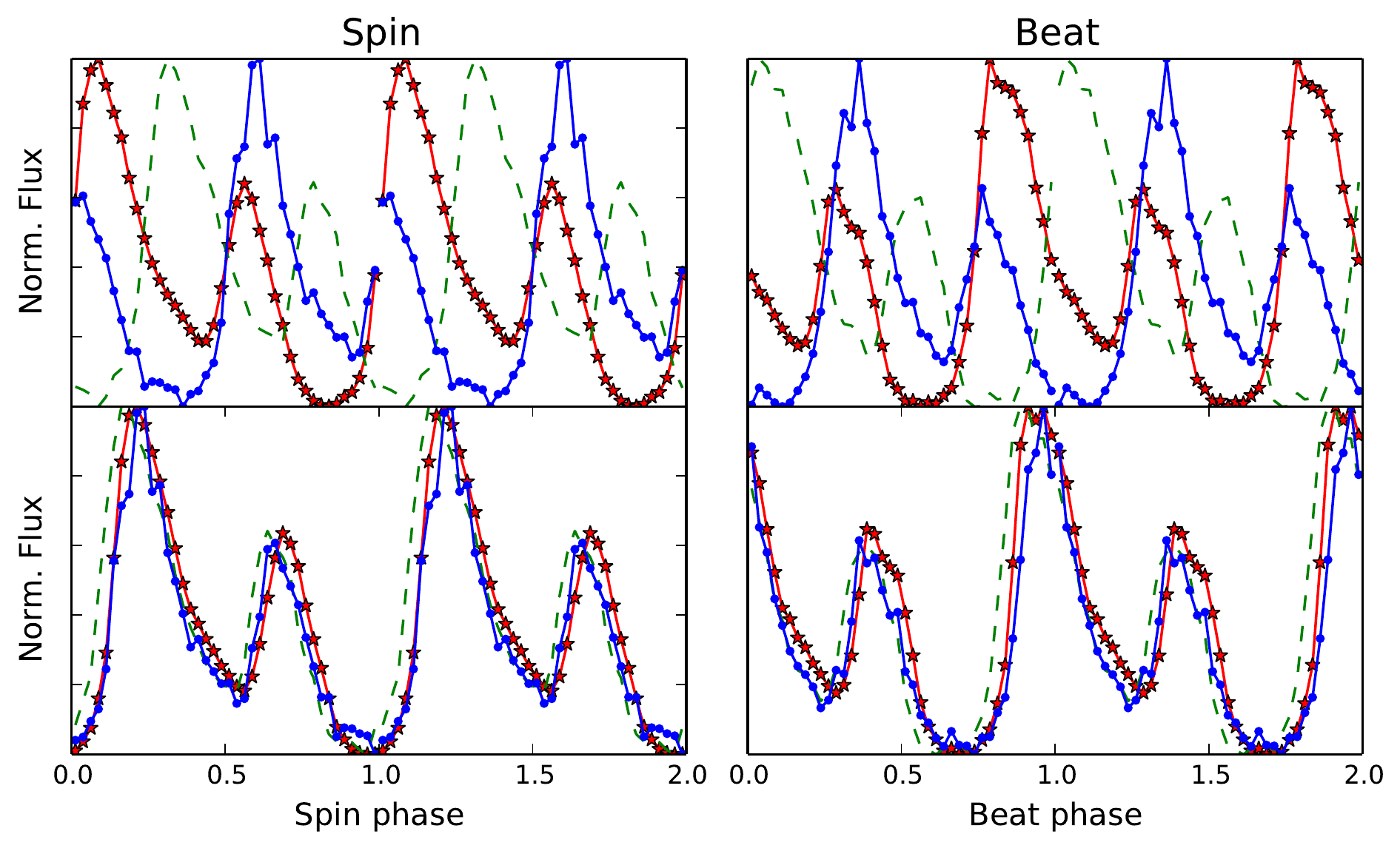}
    \caption{Left and right panels are the spin and beat phase-folded light curves. Top and bottom panels used the spin-down frequency of \citep{Marsh2016} and our new spin frequency respectively. Red, green and blue curves are the spin and beat folded light curves from the separate 2015, 2016 and 2017 data sets respectively. See text for details.}
    \label{fig:example_figure}
\end{figure}

\begin{figure}
	\includegraphics[width=\columnwidth]{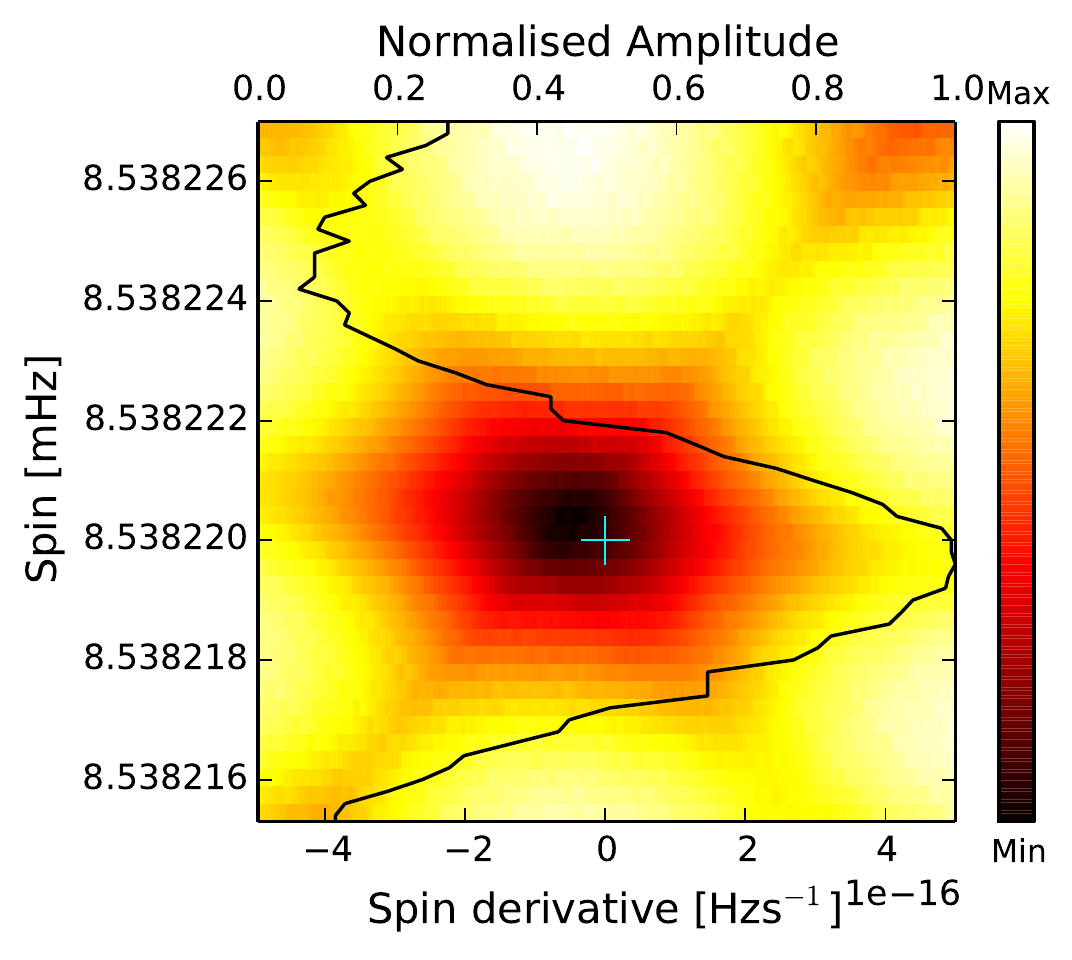}
    \caption{Cross-correlation of the three data sets (2015, 2016, 2017), spin and beat phase-folded as a function of spin and spin derivative: shown as the color map. Black pixels indicate best cross-correlation values. Black curve is the tallest peak from the total-amplitude in Fig. 5 but rotated to match the vertical spin axis. The cyan cross locates our constant spin solution from the lower 2 panels of Fig. 6. With a frequency of 8.5382346 mHz, the solution of \citep{Marsh2016} is located outside the top of the figure.}
    \label{fig:example_figure}
\end{figure}

All of the photometry, listed in Table 1, were subjected to Fourier analysis. Fig. 3 presents the amplitude spectra. We are able to identify all of the peaks in the amplitude spectra as a result of the orbital-spin sidebands and their harmonics. Clearly seen are the multiple harmonics of the spin ($\omega$) and beat ($\omega-\Omega$) frequencies as well as other sideband combinations of the spin and orbit frequencies. The spectrum is heavily aliased due to daily, weekly, monthly and yearly gaps in the dataset.

Fig. 4 shows further expanded views of the amplitude spectrum centered on the frequencies as labeled. The side-band peaks are a result of $\sim$1 year aliasing. Fig. 4 shows that the spin/beat frequencies and particularly their harmonics, calculated from the ephemeris of \cite{Marsh2016}, do not match well to the tallest peaks of the amplitude spectrum. This is further exemplified in the top 2 panels of Fig. 6 where the red, green and blue curves are the spin and beat folded light curves from the separate 2015, 2016 and 2017 data sets, respectively, where there is a clear phase offset between each successive year. The phase offset between the 2015 and 2017 observations is $\sim$0.4 in phase, which corresponds to $\sim$50-60 seconds. Therefore the ephemeris of \cite{Marsh2016} is inconsistent with our data. We stress that the ephemeris consists of three terms: a fiducial date ($T_0$), a spin frequency ($\omega$) and a spin frequency derivative ($\dot \omega$). In deriving the ephemeris all three terms are calculated simultaneously. We note that \cite{Marsh2016} do not explicitly state the fiducial date of their ephemeris, which is BMJD$_{0} = 55000.0$ expressed as a Modified Julian Day number (MJD = JD - 2400000.5) (Marsh, private communication).

Exploring our data further, Fig. 4 suggests that the tallest peaks are located at lower frequencies than suggested by the spin-down ephemeris of \cite{Marsh2016}.
However there are comparable sized peaks at several of the 1 year alias locations. Nevertheless, the amplitude spectrum in Fig. 3 shows many sidebands and harmonics of the spin and orbit frequencies. The most likely true spin (and beat) frequency, amongst the aliases, will be the one that also has corresponding high peaks amongst all the sidebands and harmonics.

Therefore, in order to eliminate between the aliases we first re-scaled the frequency axis of the magnified spectra in Fig. 4. such that they co-aligned with the fundamental of the spin frequency. For example the frequencies centered on $2\omega$ (the second right panel of Fig 4.) were divided by 2 and the frequencies centered on $3(\omega-\Omega)$ (the forth left panel of Fig 4.) had $\Omega$ added then divided by 3. The black curves in Fig. 5 show 4 examples of such spectra re-scaled with respect to the fundamental of the spin frequency. The 15 tallest amplitude peaks in Fig.3 were re-scaled in frequency and then their amplitudes  summed to produce the total amplitude spectrum, shown in red in Fig 4.

Note that each sub-spectrum was normalized by its peak amplitude before being added, otherwise the higher harmonics (lower panels) would make an insignificant contribution to the total sum despite their having good signal-to-noise ratio. Despite there being several aliases, not all of them have corresponding peaks at the various spin/orbit/beat sidebands and harmonics, resulting in a reduced summed amplitude at those frequencies. Conversely the most probable frequency has peaks in all the sub-spectra and hence has summed to produce the most significant peak in the total sum amplitude spectrum. Even without summing, it is immediately apparent from Fig. 5. that the tallest peak (amongst the aliases) in each sub  frequency spectrum are consistent with each other and therefore the most probable solution amongst the aliases.


For the above we assumed the orbital frequency of \cite{Marsh2016}. We repeated the exercise of Fig. 5 for a range of orbital frequencies (not shown). We find that the most significant orbital frequency is in agreement with \cite{Marsh2016}.

We measure the spin frequency to be $\omega=0.008538220(3)$ Hz, where the error is measured from the width of the red peak (Fig. 5) at half its amplitude. We phase-fold our observations using:
\begin{equation}
    \label{eq:ephem}
    \Phi_{t} = \mathrm{(BMJD_{t}-BMJD_{0}}) \times \omega
\end{equation}

expressed as a Modified Julian Day number (MJD = JD - 2400000.5) and BMJD$_{0} = 57530.0$ corresponding to our 2016 dataset. BMJDs are first converted to seconds.


The bottom 2 panels of Fig. 7 show the separate 2015, 2016 and 2017 data sets folded on our ephemeris and displayed in the same manner as the top panels. Compared to the top 2 panels (folded on the spin-down ephemeris of \citep{Marsh2016}), the light curves appear well aligned confirming the frequency measured from our Fourier analysis. Note that we used data between orbital phases 0.4-0.6 only, where the spin/beat pulses are the strongest and most clearly defined.

We next investigated the possibility of adding a non-zero spin-derivative to our linear ephemeris. We used a grid of spin and spin derivatives and for each grid point we produced 6 folded light curves (3 spin plus 3 beat) as in Fig. 6. The 6 light curves were then cross-correlated for each grid point.

Fig. 7 shows the results as a color coded image with the best cross-correlation values indicated as black. The tallest total-amplitude peak associated with the spin frequency (from Fig. 5.) is also over-plotted, rotated on its side. The solution for our constant frequency ephemeris is indicated by the central cyan cross.  The black region indicates that there is a range of spin and spin-derivative combinations that give comparable cross-correlation values to the zero spin-derivative solution. The black region shows that there are no spin-frequency solutions (as a result of adding a spin derivative term) that are  outside the frequency range defined by the width of the amplitude peak. However, the width (spin-derivative axis) of the black region ($\sim (-2) - (+1)\times 10^{-16}$ Hz s$^{-1}$) indicates the range of possible spin-derivatives that are currently ``hidden" in the resolution of the data. Only by adding more observations and thereby extending the time-base of the dataset can this range be narrowed.
 



We note that we assumed the orbital frequency of \cite{Marsh2016} in the cross-correlation analysis above (Figs. 6 and 7). 

\section{Summary}

We have obtained and analysed new high-speed photometric observations spanning 2 years between 2015 and 2017. Repeatable spin, beat and orbital modulations are seen throughout our dataset and are clearly defined in our Fourier analysis which displays multiple harmonics of the spin and beat pulses as well as other multiple combinations thereof.

The multiple combinations and harmonics of the spin, beat and orbital frequencies has enabled us to eliminate between aliases and to derive a sufficiently accurate spin frequency to correctly phase all of our data sets. Our spin ephemeris is inconsistent with the spin-down ephemeris reported by \cite{Marsh2016}. We suggest that the earlier observations used in the analysis of \cite{Marsh2016} are too sparse and of insufficient time resolution to derive an ephemeris as they reported. 

Currently any spin derivative within the range 
$\sim (-2) - (+1)\times 10^{-16}$ Hz s$^{-1}$ is undetectable given the time base of the observations. Longer term photometric monitoring is required in order to accurately measure the spin-evolution of the white dwarf in AR Sco essential for understanding its energetics and evolution.





\section*{Acknowledgements}

This material is based upon work by the authors which is supported financially by the National Research Foundation (NRF) of South Africa. We thank Dave Kilkenny and Encarni Romero-Colmenero for useful discussions and Tom Marsh for providing the 2015 ULTRACAM data.









\bsp	
\label{lastpage}
\end{document}